\documentclass{vgtc}                          

\ifpdf
  \pdfoutput=1\relax                   
  \usepackage{graphicx}                
  \DeclareGraphicsExtensions{.pdf,.png,.jpg,.jpeg} 
\else
  \usepackage{graphicx}                
  \DeclareGraphicsExtensions{.eps}     
\fi%

\graphicspath{{figures/}{pictures/}{images/}{./}} 

\usepackage{microtype}                 
\PassOptionsToPackage{warn}{textcomp}  
\usepackage{textcomp}                  
\usepackage{mathptmx}                  
\usepackage{times}                     
\usepackage{cite}                      
\usepackage{tabu}                      
\usepackage{booktabs}                  
\usepackage{graphicx}
\usepackage{listings}
\usepackage{color}
\usepackage{dirtytalk}

\usepackage{tikz}
\usetikzlibrary{shapes}

\usepackage{todonotes}
\usepackage{xspace}

\definecolor{dkgreen}{rgb}{0,0.6,0}
\definecolor{gray}{rgb}{0.5,0.5,0.5}
\definecolor{mauve}{rgb}{0.58,0,0.82}

\DeclareRobustCommand{\etal}{\textit{et al.}\xspace}
\DeclareRobustCommand{\bfparhead}[1]{\noindent\textbf{#1}} 

\newcommand{\orangebox}{\raisebox{0pt}{\tikz{\node[draw,scale=0.5,regular polygon, regular polygon sides=4,fill={rgb,255:red,240; green,147; blue,16},rotate=0](){};}}}

\newcommand{\greenbox}{\raisebox{0pt}{\tikz{\node[draw,scale=0.5,regular polygon, regular polygon sides=4,fill={rgb,255:red,59; green,216; blue,4},rotate=0](){};}}}

\onlineid{0}

\vgtccategory{Research}

\vgtcinsertpkg

\title{Data2Vis: Automatic Generation of Data Visualizations \\ Using Sequence to Sequence Recurrent Neural Networks}

\author{Victor Dibia \thanks{e-mail: dibiavc@us.ibm.com}\\ %
        \scriptsize IBM Research %
\and \c{C}a\u{g}atay Demiralp \thanks{e-mail: cagatay@csail.mit.edu}\\ %
     \scriptsize MIT CSAIL \& Fitnescity Labs %
     }

\abstract{
 Rapidly creating effective visualizations using expressive grammars is challenging for users who have limited time and limited skills in statistics and data visualization. Even high-level, dedicated visualization tools often require users to manually select among data attributes, decide which transformations to apply, and specify mappings between visual encoding variables and raw or transformed attributes. In this paper we introduce Data2Vis, an end-to-end trainable neural translation model for automatically generating visualizations from given datasets. We formulate visualization generation as a language translation problem where data specifications are mapped to visualization specifications in a declarative language (Vega-Lite). To this end, we train a multilayered attention-based 
encoder-decoder network with long short-term memory (LSTM) units on a corpus of visualization specifications. Qualitative results show that our model learns the vocabulary and syntax for a valid visualization specification, appropriate transformations (count, bins, mean) and how to use common data selection patterns that occur within data visualizations. Data2Vis generates visualizations that are comparable to manually-created visualizations in a fraction of the time, with potential to learn more complex visualization strategies at scale.
}

\CCScatlist{
  \CCScatTwelve{Human-centered computing}{Visu\-al\-iza\-tion}{Visu\-al\-iza\-tion techniques}{Treemaps};
  \CCScatTwelve{Human-centered computing}{Visu\-al\-iza\-tion}{Visualization design and evaluation methods}{}
}

\begin{document}



\maketitle


\section{Introduction}

\begin{figure}[t]
\centering
\includegraphics[width=\columnwidth]{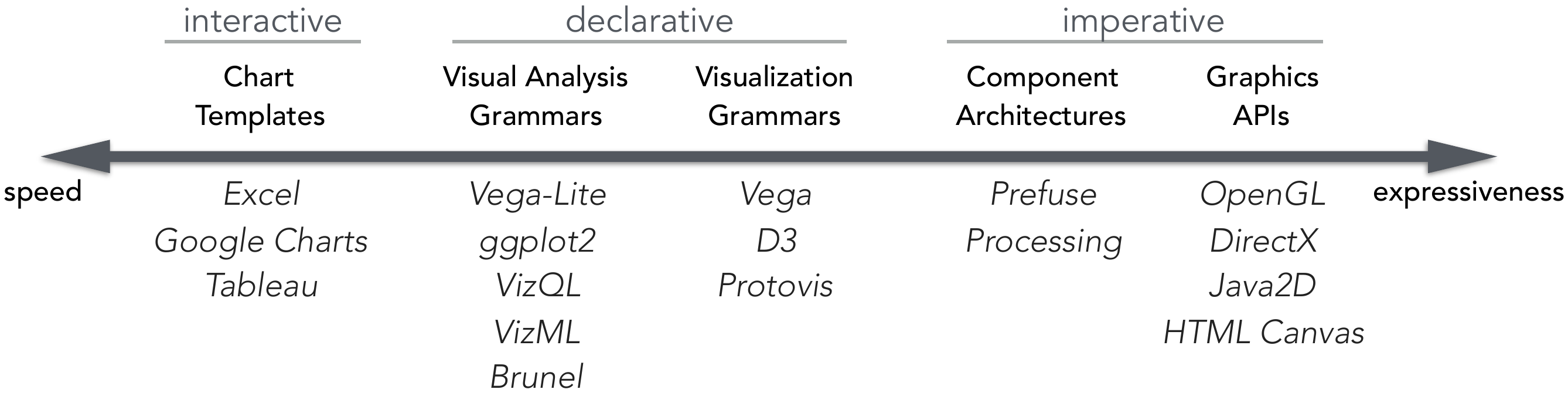}
\caption{Axis of visualization specification. Data visualizations are created
  with a spectrum of tools with a spectrum of speed and expressivity. Some of
  these tools are faster but others are more
  expressive to create visualizations.
  \label{fig:vistools_axis}
\vspace{-2.1em}
}
\end{figure}

Users create data visualizations using a range of tools with a range of
characteristics (Figure~\ref{fig:vistools_axis}).  Some of these tools are more
expressive, giving expert users more control, while others are easier to learn
and faster to create visualizations, appealing to general audiences. For
instance, imperative APIs such as OpenGL and HTML Canvas provide greater
expressivity and flexibility but require significant programming skills and
effort. On the other hand,  dedicated visual analysis tools and spreadsheet
applications (e.g., Microsoft Excel, Google Spreadsheets) provide ease of use
and speed in creating standard charts based on templates but offer limited
expressivity and customization. 

Declarative specification grammars such as ggplot2~\cite{wickham2010layered},
D3~\cite{bostock2011d3}, Vega~\cite{satyanarayan2016vega}, and
Vega-Lite~\cite{satyanarayan2017vegalite} provide a trade-off between speed and
expressivity. However, these grammars also come with steep learning curves, can
be tedious to specify depending on the syntax and abstraction level adopted,
and can suffer from reusability issues. In fact, there is little known about
the developer experience with visualization grammars, beyond the degree with which
they are used. For example, ggplot2 can be difficult for users who are not
familiar with R. Vega, which is based on a JSON schema, can be tedious even for
users who are familiar with JSON. Even tools with higher-level abstractions
such as the ones based on chart templates often require the user to manually
select among data attributes, decide which statistical computations to apply,
and specify mappings between visual encoding variables and either the raw data
or the computational summaries. This task can be daunting with complex datasets
especially for typical users who have limited time and limited skills in
statistics and data visualization. To address these challenges, researchers have proposed techniques and tools to
automate designing effective visualizations
\cite{casner:tog91,Demiralp_2014a,Key_2012,Mackinlay_1986,avd2014infovis,Roth_1994} and guide users in visual data exploration \cite{ Asimov_1985,
Demiralp:2017:VLDB, PRIM9_1974, showme:infovis07, Roth_1994, seo:infovis04,
Siddiqui_2016,Vartak_2015a,Wills_2008,Wongsuphasawat_2016,
Wongsuphasawat_2017}. 

Prior techniques and tools for automated visualization design and visualization
recommendation are based on rules and heuristics.  The need to explicitly
enumerate rules or heuristics limits the application scalability 
of these approaches and does not take advantage of expertise 
codified within existing visualizations. Automated and guided 
visualization design and exploration can significantly  benefit 
from implicitly learning these rules from examples
(i.e., data), effectively incorporating both data and visualization design
context.  

In this work, we formulate visualization design as a problem of translation between data specification and visualization specification. To operationalize our formulation, we train an LSTM-based
neural translation model (Data2Vis) on a corpus~\cite{2017-reverse-engineering-vis} of Vega-Lite visualization
specifications, taking advantage of Vega-Lite's (and of similar grammars')
design motivation to support programmatic generation. We demonstrate the
model's use in automatically generating visualizations with applications in
easing the visualization authoring process for novice users and helping
more experienced users jump start visualization design.  Our contributions
include 1) formulating visualization design as a sequence to sequence
translation problem, 2) demonstrating its viability by training a
sequence to sequence model, Data2Vis, on a relatively small training dataset
and then effectively generating visualizations of test data, and 3) integrating
Data2Vis into a web-based application that has been made publicly available at
\url{http://hci.stanford.edu/~cagatay/data2vis}. Our work is the first in
applying deep neural translation to visualization generation and has important
implications for future work, opening the way to implicitly learn visualization
design and visual analysis rules from  examples at scale. 

In what follows, we first summarize related work followed by details of the
Data2Vis model and its training process. We then present our results, providing
several visualization examples automatically generated using the trained model.
Next we discuss the potential impact of Data2Vis and its current limitations
and provide an agenda for future work. We conclude by summarizing our
contributions and insights.   
\section{Related Work}
Our work is related to earlier efforts in effective visualization
specification, automated visualization design, and deep neural networks (DNNs)
for synthesis and machine translation.    
\subsection{Declarative Visualization Specification}
Earlier data visualization work proposes grammars and algebraic operators over
data as well as visual encoding and design variables to specify visualizations 
(Figure~\ref{fig:vistools_axis}).
Wilkinson's seminal work~\cite{wilkinson:book99} introduces  a grammar of
graphics and its implementation (VizML), greatly shaping the subsequent research
on visualization specification. Polaris~\cite{Stolte_2002} (commercialized as 
Tableau) uses a table algebra drawn from Wilkinson's grammar of graphics.  The
table algebra of Polaris later evolved to VizQL~\cite{hanrahan2006vizql},
forming the underlying representation of Tableau visualizations. Wickham
introduces ggplot2~\cite{wickham2010layered}, a widely-popular package in the R
statistical language,  based on Wilkinson's grammar.  Similarly,
Protovis~\cite{bostock2009protovis}, D3~\cite{bostock2011d3},
Vega~\cite{satyanarayan2016vega}, Brunel~\cite{wills2017brunel}, and
Vega-Lite~\cite{satyanarayan2017vegalite} all provide grammars to declaratively
specify visualizations. Some of them require more complete specifications than
others. For instance, Protovis, D3 and Vega support finer control over
visualization specification with incurred cost of verbosity.  

\begin{figure}[t]
  \centering
  \includegraphics[width=\columnwidth]{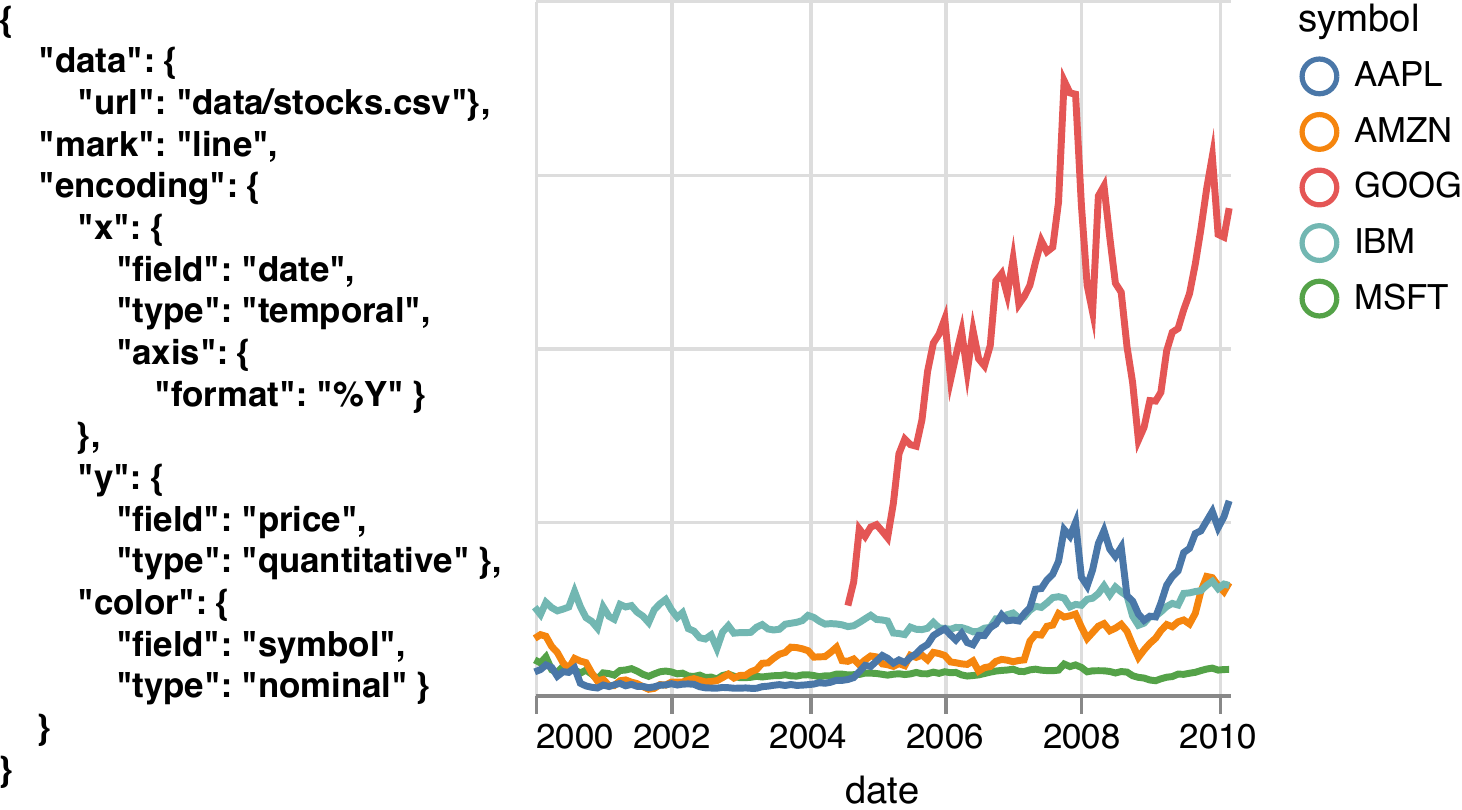}
  \caption{A Vega-Lite specification (left) and the generated visualization
    (right). Users can succinctly specify selections, transformations
    and interactions using the Vega-Lite grammar formatted in JSON
    ~\protect\cite{satyanarayan2017vegalite}.
    \label{fig:vegaliteinterface}} 
\end{figure}

Wongsuphasawat et al.~\cite{Wongsuphasawat_2016} introduce Vega-Lite
(Figure~\ref{fig:vegaliteinterface})  to support Voyager, a faceted browser for
visualization recommendations. Vega-Lite is a high-level grammar built on top
of Vega  to facilitate  clarity  and conciseness with some loss in
expressivity. The expressivity of Vega-Lite is a strict subset of Vega. We
train our model on a Vega-Lite corpus~\cite{2017-reverse-engineering-vis},
which contains datasets and corresponding visualizations specified in
Vega-Lite.  

Declarative grammars eschew chart templates typically used in dedicated
visualization tools or spreadsheet applications such as Microsoft Excel and
Google Spreadsheets, which have limited support for customization.  Conversely,
these grammars facilitate expressivity by enabling a combinatorial composition
of low-level building blocks such as graphical marks, scales, visual encoding
variables, and guides. However, increased expressivity often decreases 
the speed with which visualizations can be created and makes the learning
more difficult, limiting the number of users who can effectively use the
specification method.  One of our aims with Data2Vis is to bridge this gap
between the speed and expressivity in specifying visualizations.  

\subsection{Automated Visualization}
Prior work proposes desiderata and tools (e.g., \cite{casner:tog91,
Demiralp_2014a, avd2014infovis,Mackinlay_1986,Roth_1994}) to automatically
design effective visualizations, building on Bertin's study~\cite{Bertin_1983}
of visual encoding variables and earlier graphical perception research, e.g.,
~\cite{maceachren:book95,
shortridge:ac82,cleveland:jasa84,lewandowsky:jasa89,Mackinlay_1986,
shepard:sci87}. Earlier research also develops interactive systems and
recommendation
schemes~\cite{Bouali_2016,Gotz_2008,showme:infovis07,seo:infovis04,
Siddiqui_2016, topk2017sigmod, Vartak_2015b, Vartak_2015a,
scagnostics2005infovis, Wills_2008, Wongsuphasawat_2016, Wongsuphasawat_2017}
to guide users in exploratory data analysis and visualization design.
PRIM-9~\cite{PRIM9_1974}, GrandTour~\cite{Asimov_1985}
SeeDB~\cite{Vartak_2015a}, Zenvisage~\cite{Siddiqui_2016},
ShowMe~\cite{showme:infovis07}, Voyager~\cite{Wongsuphasawat_2016}, Voyager
2~\cite{Wongsuphasawat_2017}, SAGE~\cite{Roth_1994} and VizDeck~\cite{Key_2012}
prioritize charts according to one or more evaluation measures such as data
saliency, data coverage, perceptual effectiveness, user task, and user
preferences.  Similarly, Rank-by-Feature~\cite{seo:infovis04},
AutoVis~\cite{Wills_2008}, and Foresight~\cite{Demiralp:2017:VLDB} use
statistical criteria over data attributes and instances in recommending and
ranking visualizations.

Data2Vis represents a departure from rule-based approaches of prior work both
in conceptual formulation and technical approach taken. It makes contributions
by specifying how automated visualization can be cast as a learning problem,
providing a concrete implementation of a deep learning model for visualization
generation. Data2Vis emphasizes the creation of visualizations specifications
using rules \textit{learned} from examples, without resorting to a predefined
enumeration of rules or heuristics, complementing earlier work.  Researchers
recently recognized the potential of machine learning in automating
visualization design and visual analysis~\cite{Saket:2018:Learning}, applying
machine learning for recommending visualizations~\cite{Saket:2018:TVCG,
Hu:2018:VizML, Yuyu:2018:DeepEye} and refining visualization
recommendations~\cite{Moritz:2019:Draco}.  Data2Vis differs from this exciting
line of recent work, which relies on feature extraction and manual constraint
specification, in learning to automatically generate visualizations from data
with an end-to-end approach. 

Adopting a learning  approach to designing automated visualization systems
holds potential for improving the maintenance and scalability of such systems.
Existing approaches are limited by a dependence  on a set of manually created
(interdependent) rules which can be voluminous, tedious update, and may not
sufficiently cover edge cases necessary to generate good visualizations. By
using a learning approach, we avoid these limitations as a learned model can
better represent the visualization rule space given sufficient examples.
Further more, the performance and capabilities of the system can be improved by
improving the dataset of examples used to train models within learning based
systems. As more users author visualizations, the system  can leverage
\textit{experiences and rules} encoded within these visualizations, to increase
it's coverage and  \textit{scale} its performance. The visualization generation
capabilities of Data2Vis can also be integrated into existing higher-level
recommendation systems of visual data exploration and used in tandem with
rule-based techniques to drive these systems.  We published the current work
earlier as a preprint~\cite{ Dibia:2018:Data2Vis} and made the source code for
the Data2Vis  model publicly available~\cite{Data2Vis:2018:GitHub}.

\subsection{Deep Neural Networks for Synthesis}
Prior deep neural network (DNN) research studies adopt generative approaches
to learn human-like cognitive and creative capabilities.  Examples include the
use of models to synthesize music, drawings, images from textual descriptions,
code from hand-drawn sketches or interface screenshots.  Ha et
al.~\cite{HaSketchRNNE17} train a recurrent neural network (RNN) to predict and
generate stroke-based drawings of common objects.  Reed et al.
~\cite{reed2016generative} present a DNN architecture and generative adversarial 
network (GAN) formulation to ``translate'' textual visual concepts to pixels.  
Others learn how to generate code from  user interface
screenshots~\cite{beltramelli2017pix2code} and how to compose music using
purely sequential models~\cite{eck2002first,johnson2017generating} and
cascading a sequential model with  a restricted Boltzman
machine~\cite{boulanger2012modeling}.  All these approaches aim to simplify the
creative process for both novices and experts. In this sense, our work here
shares a motivation with prior work. We also use a variation of
sequential neural network models, a sequence to sequence model, 
to generate visualization specifications from given data.  

\begin{figure*}[t]
  \includegraphics[width=\textwidth]{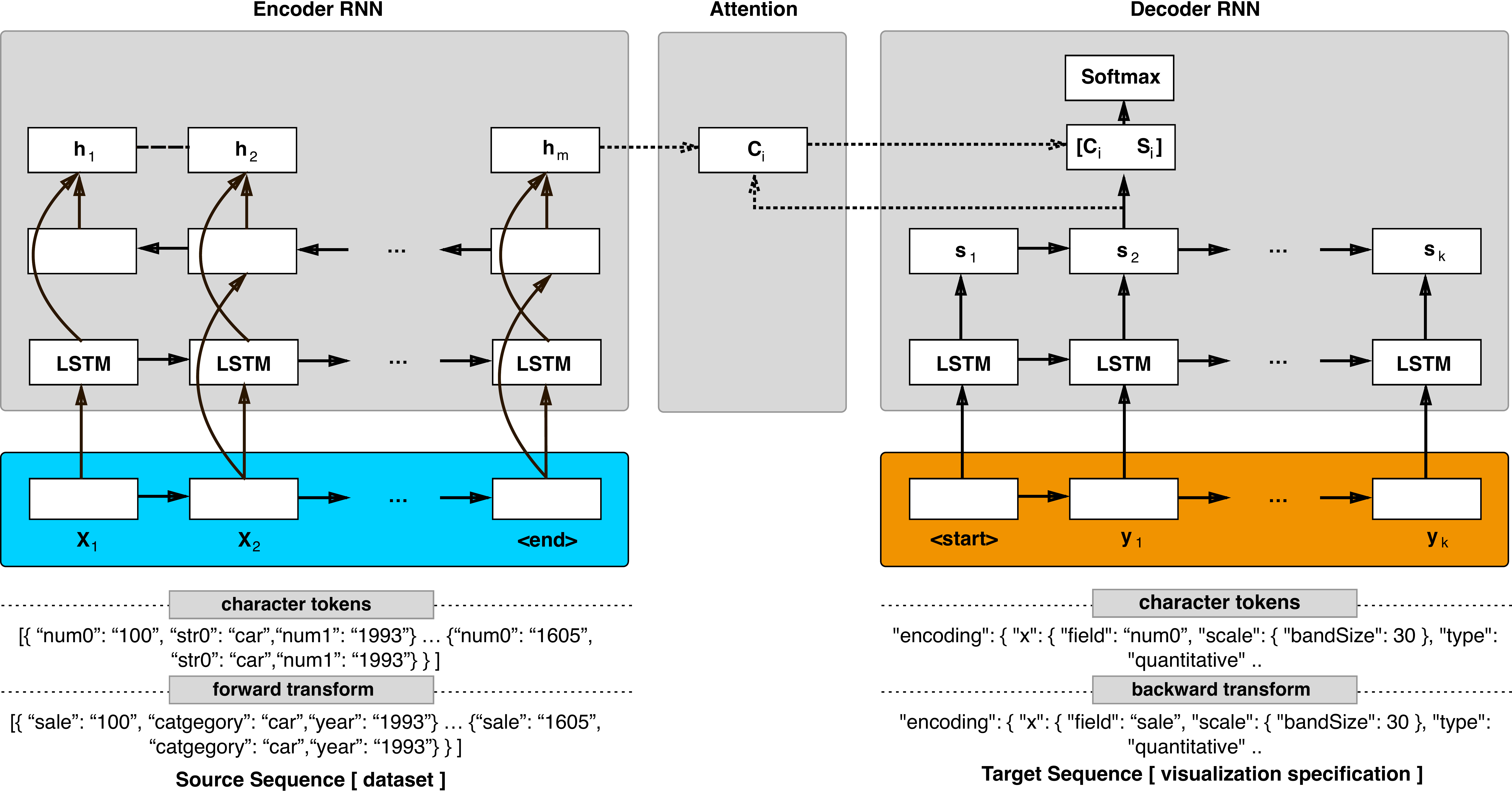}
  \caption{Data2Vis is a sequence to sequence model with encoder-decoder
  architecture and attention module. To simplify learning, we perform simple forward and backward transformations on the source (dataset in JSON format) and target sequence (Vega-Lite visualization specification) which are then converted to character tokens.}
  \label{fig:modeldiagram}
\end{figure*}

\subsection{Deep Neural Networks for Machine Translation} Recent work
introduces DNN models, e.g., ~\cite{Bahdanau2014, ChoMGBSB14,
kalchbrenner2013recurrent, Luong2015, SutskeverVL14} that significantly
improves~\cite{jean2014using,
luong2014addressing,sennrich2016edinburgh,wu2016google} the performance of
machine translation systems, surpassing the preceding phrase-based approaches.
Deep neural translation models eschew hand engineering the features, in large
part, by using large training data, enabling the end-to-end learning in
practice. Sequence to sequence models (e.g.,~\cite{Bahdanau2014,Luong2015}) are
a particularly successful and popular  class of deep learning models applied in
machine translation (see ~\cite{britz2017architectures} for an evaluation of
alternative architectures). Akin to autoencoders, these models have also a
symmetric, encoder-decoder architecture.  Sequence to sequence models are
composed of encoder-decoder layers which consist of recurrent neural networks
(RNNs) and an attention mechanism that aligns target tokens with source tokens.

In addition to translating between natural languages, earlier work, e.g.,~\cite{BalogGBNT16,ChenLS18,DevlinUBSMK17,LingGHKSWB16,ParisottoMSLZK16,YinN17} also uses DNNs to translate between two domain specific languages (DSLs), 
between a natural language specification and a DSL (e.g. translating from natural language to SQL \cite{Dong2016LanguageAttention, Zhong2017Seq2SQLLearning}), and 
between two programming languages. Similar to the prior work translating between general or domain specific programming languages, Data2Vis also translates between  two formal languages. 
Ling et al.~\cite{LingGHKSWB16} use a sequence to sequence model to translate TCG (Trading Card Games) cards to their Python and Java specifications without
explicitly representing the target syntax. Data2Vis is also a sequence to sequence model that directly uses textual source and target specifications without representing their syntax (e.g., using abstract syntax trees) explicitly.  

\begin{figure}[h] \centering
  \includegraphics[width=\columnwidth]{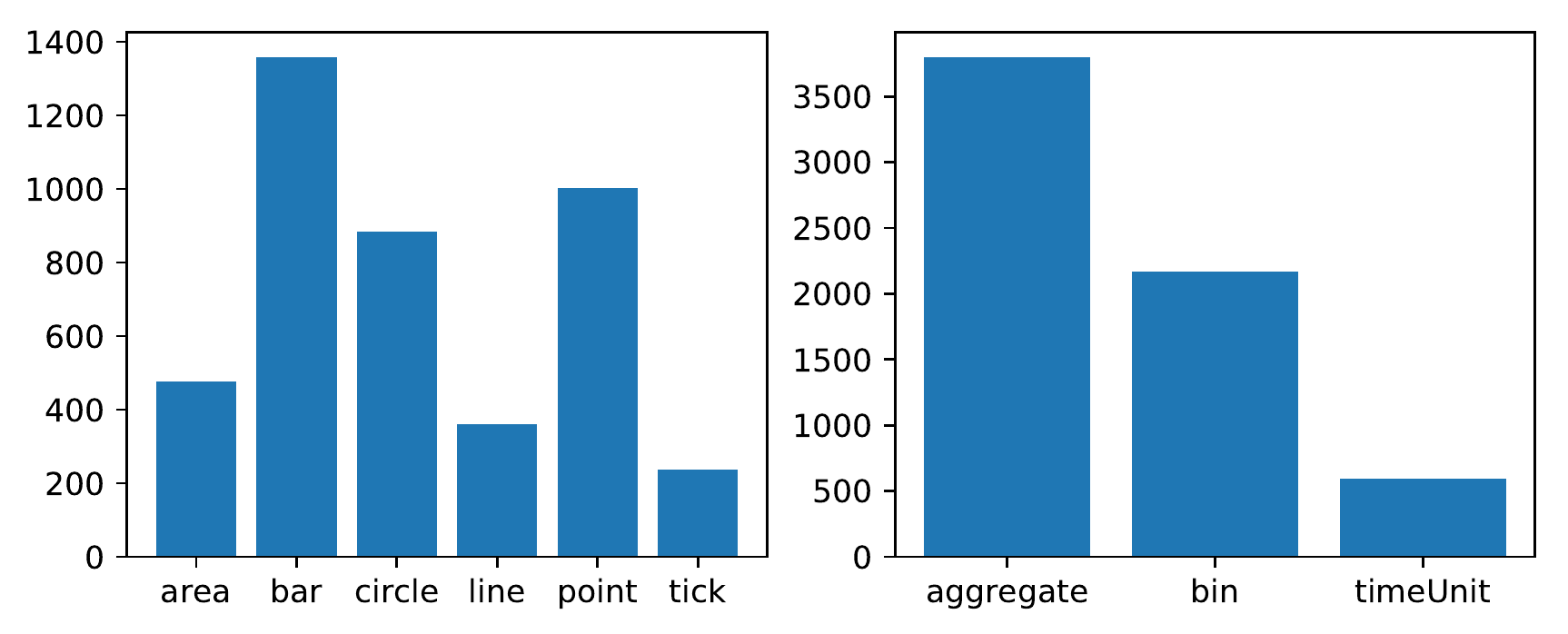}
  \caption{Frequency of the Vega-Lite mark types and transforms used in 
    our training examples.\label{fig:datasetcharacteristics}} 
\end{figure}

\section{Problem Formulation } 
Building on earlier work  that applies deep learning for translation and
synthesis, we formulate the data visualization problem as a sequence to sequence translation problem, which can be readily addressed using sequence to sequence models (seq2seq)~\cite{Bahdanau2014,ChoMGBSB14,SutskeverVL14}. Our input sequence is a dataset (fields, values in json format) and our output sequence is a valid Vega-Lite \cite{satyanarayan2017vegalite,satyanarayan2016vega} visualization specification.

Existing models used for sequence translation \cite{ChoMGBSB14, SutskeverVL14, Bahdanau2014,britz2017architectures,Luong2016,Luong2015}  belong to a family of \textit{encoder-decoder} networks where the \textit{encoder} reads and encodes a source sequence into a fixed length vector, and a \textit{decoder} outputs a translation based on this vector. The entire encoder-decoder system is then jointly trained to maximize the probability of outputting a correct translation, given a source sequence.

While sequence to sequence models have originally focused on generating data that is sequential or temporally dependent e.g language translation \cite{ChoMGBSB14, SutskeverVL14, Bahdanau2014,britz2017architectures,Luong2016,Luong2015}, they also find applications for problems where the output or input is non-sequential as seen in text summarization \cite{Nallapati2016,SumitChopra2016} and image captioning \cite{karpathy2015deep,vinyals2015show,xu2015show}. Two important advances that enable non-sequential use cases include the introduction of bidirectional RNN units \cite{Schuster1997} and attention mechanisms \cite{xu2015show,vinyals2015show,Bahdanau2014}. 
An ordinary RNN (unidirectional) reads an input sequence $x$ from the frist token $x_1$ to the last $x_{m}$ and generates an encoding only based on the preceding tokens it has seen. On the other hand, a Bidirectional RNN (BiRNN) consists of both a forward RNN and a backward RNN, which enables an encoding generation based on both the preceding and following tokens. The forward RNN \overrightarrow{f} reads the input sequence as it is ordered (from $x_1$ to $x_{m}$) and calculates a sequence of forward hidden states ( \overrightarrow{ h_1}, ... ,\overleftarrow{h_{m}} ). 
The backward RNN \overleftarrow{f} reads the sequence in the reverse order (from $x_{m}$ to $x_1$), resulting in a sequence of backward hidden states 
( \overleftarrow{ h_1}, ... ,\overleftarrow{h_{m}} ). Thus, when a BiRNN is used to encode an input sequence, it generates a hidden state \overrightarrow{h_j} which is a concatenation of both the forward and backward RNNs, $h_j = \Big[ \overrightarrow{ h_j}^\top ; \overleftarrow{ h_j}^\top \Big]^\top $ and contains summaries of both the preceeding and following tokens.
Attention mechanisms allow a model to focus on aspects of an input sequence while generating output tokens. They provide the additional benefits of making translation models robust to performance degradation while generating lengthy sequences, and enable the model to learn mappings between source and target sequences of different lengths \cite{Bahdanau2014}. For example, when used in image captioning, attention mechanisms allow the model to focus on specific parts of objects in an image, while generating each word or token in the image caption. Furthermore, attention mechanisms improve our ability to interpret and debug sequence to sequence models as they provide valuable insights on \textit{why} a given token is generated at each step. Taken together, these two important advances enable us to use a sequence translation model that first takes into consideration the entire data input (dataset) and then focus on aspects of the input (fields) in generating a visualization specification.

Seq2seq models for language translation are trained using embeddings of the
source and target tokens which can be generated based on words, subword or per character units~\cite{Bahdanau2014,ChoMGBSB14,SutskeverVL14}.  We select a per character unit tokenization given our source and target sequences consist of symbols as opposed to learnable word groups seen in related problems like language translation. 


\section{Model}
Our model (Figure~\ref{fig:modeldiagram}) is based on an encoder-decoder
architecture with attention mechanism that  has been previously applied in
machine translation ~\cite{Bahdanau2014,Luong2016,Luong2015}. The encoder
is a bidirectional recurrent neural network (RNN) that takes in an input
sequence of source tokens \( x = (x_1, ... , x_m) \) and outputs a sequence
of states \( h = (h_1, ... , h_m) \). The decoder is also an RNN that
computes the probability of a target sequence \( y = (y_1, ... , y_k) \)
based on the hidden state $h$. The probability of each token in the target
sequence is generated based on the recurrent state of the decoder RNN,
previous tokens in the target sequence and a context vector \( c_i \). The
context vector (also called the attention vector) is a weighted average of
the source states and designed to capture the context of source sequence
that help predict the current target token. 

We use a 2-layer bidirectional RNN encoder and a 2-layer RNN decoder, each
with 512 Long Short-Term Memory
(LSTM)~\cite{learningto99gers,lstm97hochreiter} units (cells). To decide
which RNN unit type to use, we experimented with both gated recurrent unit
(GRU)~\cite{ChoMGBSB14} and LSTM, both of which are common RNN cell variants.
We found LSTM cells provided better results (valid json, valid Vega-Lite specification) compared to GRU cells, which concurs
with earlier empirical results~\cite{britz2017architectures}.

\section{Data and Preprocessing}
To generate plausible visualizations conditioned on a given source dataset, our
model should achieve several learning objectives.  First, the model must select
a subset of fields to focus on when creating visualizations (most datasets have
multiple fields which cannot all be simultaneously visualized). Next, the model
must learn differences in data types across the data fields (numeric,
string, temporal, ordinal, categorical etc.), which in turn guides how each
field is specified in the generation of a visualization specification. Finally,
the model must learn the appropriate transformations to apply to a field given
its data type (e.g., aggregate transform does not apply to string fields). In
our case, this includes view-level transforms (aggregate, bin, calculate,
filter, timeUnit) and field level transforms (aggregate, bin, sort,  timeUnit)
supported by the Vega-Lite grammar.

Achieving these objectives using a character based sequence model can be
resource intensive. While character based models result in smaller
vocabulary size and are more accurate for specialized domains, they also
present challenges --- a character tokenization strategy requires more
units to represent a sequence and requires a large amount of hidden layers
as well as parameters to model long term dependencies~\cite{BojanowskiJM15}.  To address this issue and scaffold the learning
process, we perform a set of transformations. First, we replace string and
numeric field names using a short notation --- ``str'' and ``num''  in the
source sequence (dataset). Next, a similar backward transformation (post
processing) is eplicated in the target sequence to maintain consistency
in field names (see Figure~\ref{fig:modeldiagram}).  These
transformations help scaffold the learning process by reducing the
vocabulary size, and prevents the LSTM from learning field names (as we
observed in early experiments). In turn we are able to reduce the overall
source and target sequence length, reduce training time and reduce the
number of hidden layers which the model needs to converge. Our training dataset is constructed from 4300 Vega-Lite visualizations
examples, based on 11 distinct datasets. 
The examples were originally compiled
by~\cite{2017-reverse-engineering-vis} where the authors use the CompassQL \cite{Wongsuphasawat_2016CompassQL} recommendation engine within Voyager2 ~\cite{Wongsuphasawat_2016} to generate charts with 1-3 variables, filtered to remove problematic instances. These charts are generated based on heuristics and rules which enumerate, cluster and rank visualizations according to data properties and perceptual principles ~\cite{Wongsuphasawat_2016}. While these examples contain a simplified range of transformations and do not encode any interactions, they  represent valid Vega-Lite examples and conform to important perceptual principles enforced by rules within Voyager2. These characteristics make the dataset a suitable, low-complexity test bed for benchmarking  our model's performance on the task of learning to generate visualizations given only input data.  




Similar to datasets observed in the wild, our sample dataset contains  charts with 6 different types of visualizations (area, bar, circle, line, point, tick) and
three different transforms (aggregate, bin, timeUnit)(see
Figure~\ref{fig:datasetcharacteristics}). Based on this similarity, we expect similar learning performance when our model is trained with real world data sets.  
To generate our training dataset, we iteratively generate a source (a single row from the dataset) and target pair (see Figure~\ref{fig:modeldiagram}) from each example file. Each example is then sampled 50 times (50 different data rows with the same Vega-Lite specification) resulting in a total of 215,000 pairs which are then used to train our model. 


\subsection{Training}
We begin by generating a character vocabulary for our source and target
sequences (84 and 45 symbols, respectively).  A dropout rate of 0.5 is
applied at the input of each cell and a maximum source and target sequence
length of 500 is used. The entire model is then trained end-to-end using a
fixed learning rate of 0.0001 with Adam optimizer, minimizing the negative
log likelihood of the target characters using stochastic gradient descent.
Our implementation is adapted from an open source neural machine translation framework by Britz \etal~\cite{britz2017architectures}. We
train our model for a total of 20,000 steps, using a batch size of 32. We achieve a translation performance log perplexity metric score of 0.032, which suggests the model excels at predicting visualization specifications that are similar to specifications in our test set.


    

\begin{figure}[h] \centering
  \includegraphics[width=\columnwidth]{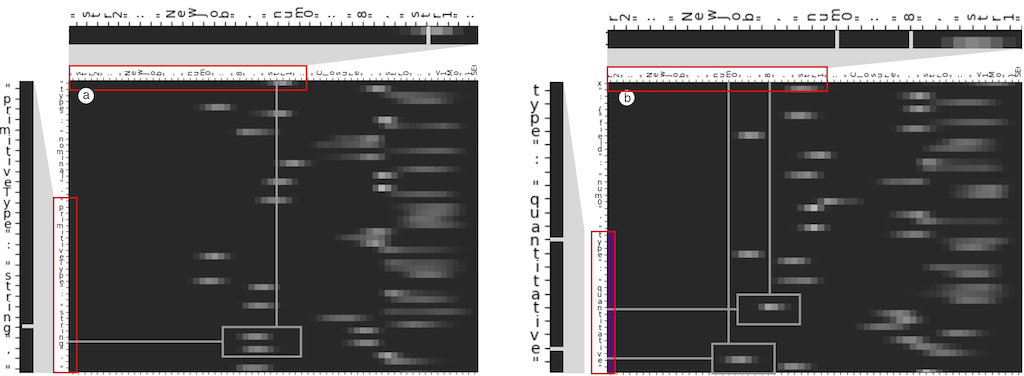}
  \caption{Example attention plots for a visualization generation case (a) Model learns to pay attention to field name "str" in generating the "string" field type applied to the field. (b) Model learns to pay attention to the field name "num0" and its value in specifying the "quantitative" field type applied to the field.\label{fig:attentionplot}} 
\end{figure}
\begin{figure*}[t]
  \includegraphics[width=1\textwidth]{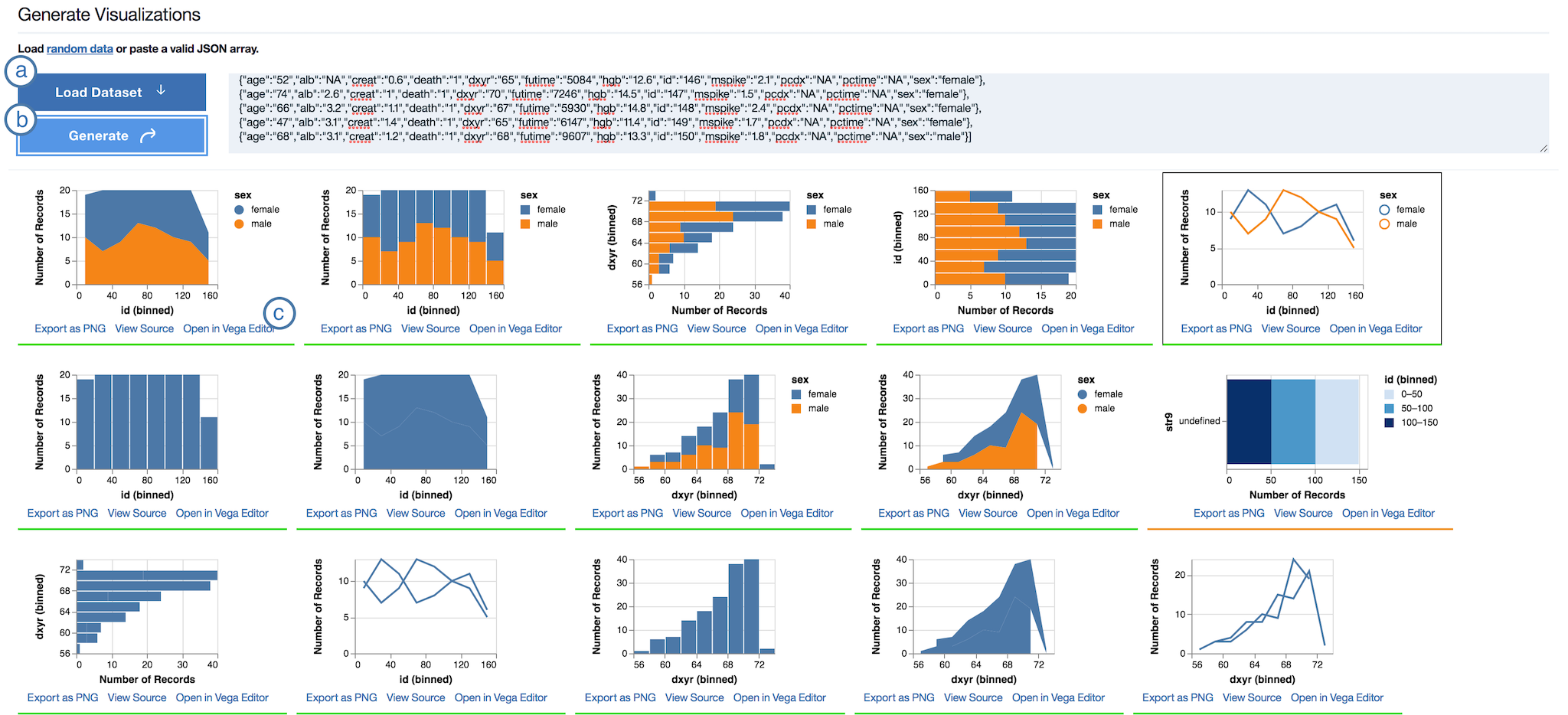}
  \caption{Data2Vis qualitative evaluation interface with results from beam search. (a) A user can load a random dataset from the RDdataset collection or paste a dataset (JSON format) and select ``Generate.'' (b) User can paste a JSON dataset and select "Generate"(c) Data2Vis generates Vega-Lite specifications using beam search (beam width = 15 in this case) based on the dataset. The user can modify and iterate on any of the visualizations using the Vega-Lite editor.  Highlights below each visualization represents cases of valid specifications \protect\greenbox\xspace and incomplete specifications \protect\orangebox\xspace where the model attempts to use variables not in the dataset (phantom variables).
    \label{fig:vizinterface}}
\end{figure*}

\begin{figure*}[ht] 
  \includegraphics[width=\textwidth]{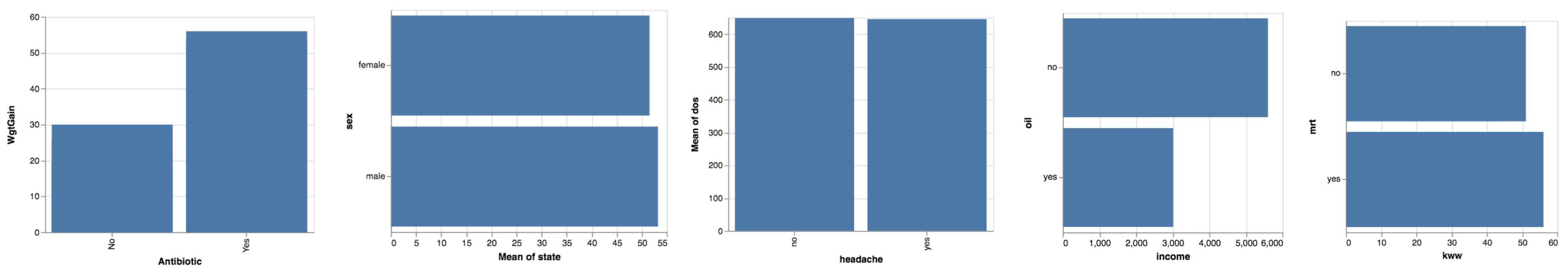}
  \caption{Examples of visualizations where the model has learned common selection patterns and leverages concepts such as responses (yes, no) 
  and sex (male, female).\label{fig:multivariatesubset}}
\end{figure*}

\begin{figure*}[ht]
  \includegraphics[width=\textwidth]{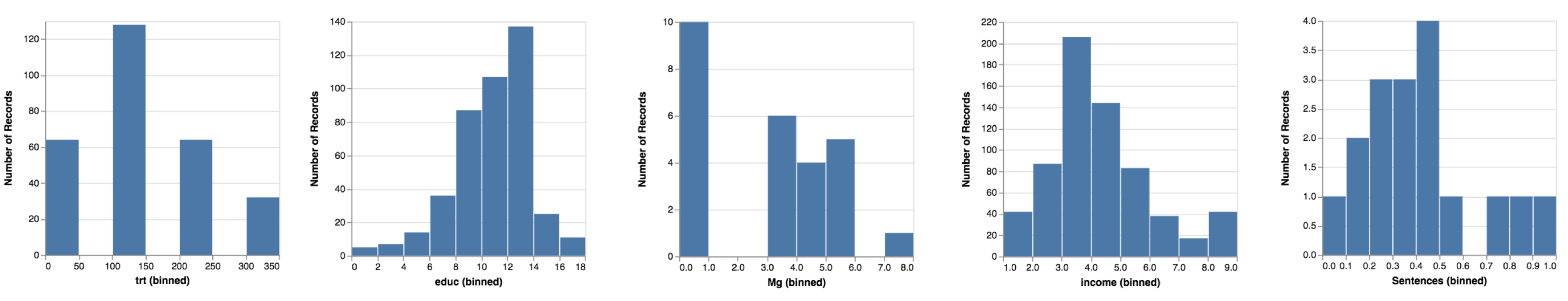}
  \caption{Examples where the model has learned to generate univariate plots that summarize 
    fields selected from the dataset.\label{fig:univariateplots}}
\end{figure*}

\begin{figure*}[ht] 
  \includegraphics[width=\textwidth]{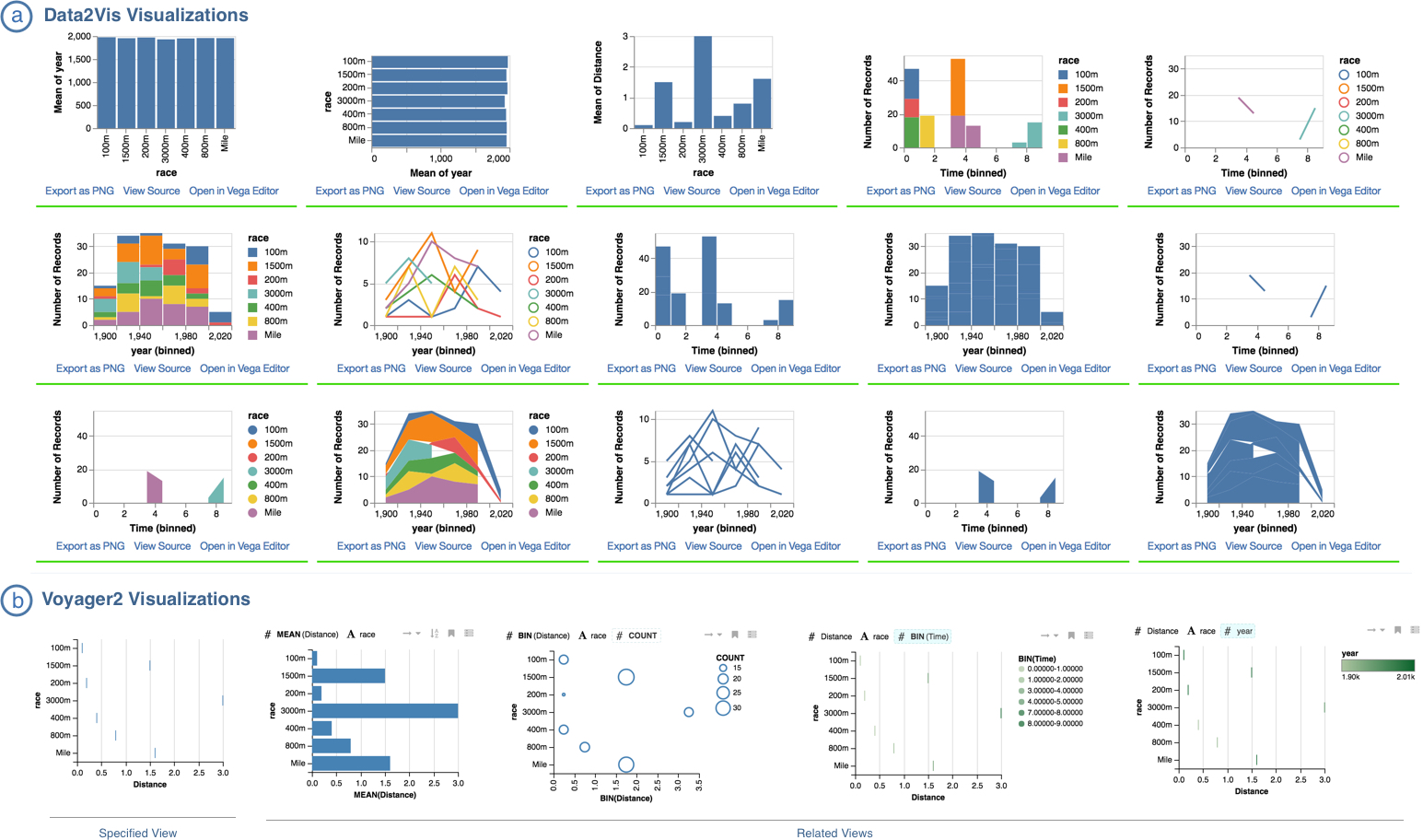}
  \caption{A comparison of visualizations generated by Data2vis(a) and Voyager~2(b) given 
    the same race dataset.
  \label{fig:comparison}} 
\end{figure*}

\section{Results}

\subsection{Examples of Automated Visualization Generation} Quantifying the
performance of a generative model can be challenging. Following existing
literature
~\cite{johnson2017generating,KarpathyJL15,DBLP:journals/corr/abs-1710-10196},
we explore a qualitative evaluation of the model's output. To evaluate the
model, we use the Rdataset repository \footnote{Rdatasets is a collection of
1147 datasets originally distributed alongside the statistical software
environment R and some of its add-on packages.} (cleaned and converted to a
valid JSON format) which was not included in our training.
Figure~\ref{fig:vizinterface} shows visualizations generated from a randomly
selected dataset in the Rdataset collection.  The range of valid univariate and
multivariate visualizations produced suggests the model captures aspects of the
visualization generation process. As training  progresses, the model
incrementally learns the vocabulary and syntax for valid Vega-Lite
specifications, learning to use quotes, brackets, symbols and keywords.  The
model also appears to have learned to use the right type of variable
specifications in the Vega-Lite grammar (e.g. it correctly assigns a string
type for text fields and a quantitative for numeric fields). Qualitative
results also suggest the use of appropriate transformations (bins, aggregate)
on appropriate fields (e.g. means are performed on numeric fields). The model
also learns about common data selection patterns that occur within
visualizations and their combination with other variables to create bivariate
plots. As experts create visualizations, it is common to group data by
geography (country, state, sex), characteristics of individuals (citizenship
status, marital status, sex) etc. Early results suggests that our model begins
to learn these patterns and apply them in its generation of visualizations. For
example, it learns to subset data using common ordinal fields such as responses
(yes/no), sex (male/female) etc and plots these values against other fields
(Figure \ref{fig:multivariatesubset} ).  Finally, in all cases, the model
generates a perfectly valid JSON file and valid Vega-Lite specification with
some minor failure cases (Figure
\ref{fig:vizinterface}\protect\orangebox\xspace).



\subsection{Beam Search} To explore a variety of generated visualizations, a
simple beam search decoding algorithm described in Wu et
al.~\cite{wu2016google} is used. As opposed to outputting the most likely
(highest probability) translation of an input sequence, beam search expands all
possible next steps during generation and keeps the \textit{k} most likely,
where k is a user specified parameter known as the \textit{beam width}. Unlike
conventional language translation systems where beam search is applied mainly
to improve translation quality by maximizing conditional probabilities of
generated sequences \cite{Graves2012}, we also
explore beam search as a way to generate a \textit{diverse} set of candidate
visualizations by outputting all parallel beam results. With beam search, we
observe the model generates more diverse plots, exploring combinations of chart
types and the use of multiple variables. Figure \ref{fig:vizinterface} shows
results from beam search (beam width=15) where the model focuses on two fields
from the dataset, generates univariate plots for these fields, subsets the
plots by sex (male/female) and uses three chart types (bar, area, line).


\subsection{Attention Plots} To further explore the efficacy of our model, and
ascertain how well it learns to use aspects of the input data in generating
visualizations, we examine plots of the attention weights (Figure
\ref{fig:attentionplot}) assigned to each predicted token.  Results suggest
that the model assigns non-monotonic weights to different input characters
while generating the parts of the specification such as the fields used for
each visualization axis, the data types assigned and the transformations
applied to each field. As shown in Figure \ref{fig:attentionplot}, the model
places strong weights on the characters ``num0'' and its value ``8'' while
generating the "quantitative" data type which it has assigned to an axis.

\subsection{Comparison with a Visualization Recommender}

We compare results from Data2Vis with results from
Voyager~2~\cite{Wongsuphasawat_2017}. Note that while Voyager~2 recommends
visualizations, it requires the user to select a set of data fields of interest
(limited to two selections) and additional preferences. Thus, for the purpose
of qualitative comparison, we present both tools with the same
\href{https://github.com/vincentarelbundock/Rdatasets/blob/master/csv/DAAG/progression.csv}{race
progression}  dataset, and select two fields to view recommendations from
Voyager~2. Qualitative results are presented in Figure \ref{fig:comparison}
which demonstrate that Data2Vis generates a richer variety of charts.
Visualizations generated by Data2Vis are not limited to specific constraints,
demonstrating its viability for the task of generating a manageable set of
visualizations based on data. 

\subsection{Web Application Integrating Data2Vis}  To further evaluate the
utility of our model, we developed a web application prototype interface
(Figure \ref{fig:vizinterface}) that supports the use case of an analyst
exploring data similar to ~\cite{Wongsuphasawat_2016,Wongsuphasawat_2017}. The
interface supports three primary interactions; data import, visualization
generation and visualization update. First, the analyst is able to import a
dataset into the application.  They can do this by using the ``load dataset''
button which loads  a randomly selected dataset from the Rdataset repository or
they can directly paste a JSON data array in the provided text field. Next, the
analyst can select the ``generate'' button which submits the dataset to our
model, receives a Vega-Lite specification (placed in a text field) and renders
the plot. Finally, the analyst can update the generated specification by
opening it in the Vega-Lite editor. We showed this early prototype to two
visualization researchers and our observations suggest they were able to
quickly build on the specifications generated by the model, making changes to
field selections and transformations.  

\section{Discussion}

We presented  the very first attempt to transform data to visualizations using
a deep neural network and apply a neural machine translation (seq2seq) model in
automating visualization generation. Below, we discuss the potential impact of
our approach and limitations of the current Data2Vis model along with future
research directions. 

\subsection{Impact and Use Case}
\bfparhead{Making Visualization Authoring Easier} Providing users with little
or no programming experience with the ability to rapidly create expressive data
visualizations empowers users and brings data visualization into their personal
workflow. Based on our early findings, Data2Vis is able to learn patterns in data visualizations that are can be generalized to a variety of real world datasets. For example, the use of categorical variables like gender, sex, location (state, country) and the tendency to focus on such variables can be learned from data. Thus, visualizations generated which encode such principles holds potential to make data visualization more accessible, speed-up the visualization authoring process and augment the visualization capabilities of all users.


\bfparhead{Accelerating Data Exploration}
For visualization experts, it is likely that visualizations created by Data2Vis
may be insufficient for their needs. This is especially true when the structure
of the data being analyzed is unknown or unusual and effective exploration of
the data requires complex transforms as well as deep domain expertise. However,
Data2Vis can contribute to this process by ``jumpstarting'' the
visualization process---first by generating a set of valid visualization specifications and \textit{seeding} the creativity process with these initial visualizations. Analysts can initialize their visualization tasks with Data2Vis and iteratively \textit{correct} its content while generating intermediate visualizations.  


\subsection{Limitations} 

\bfparhead{Field Selection and Transformation}
The current version of our model has limitations which occur in about
15-20\% of tests. First, the model occasionally selects what we refer to as a
phantom field (a field that does not exist in the input dataset) as part of the
visualization spec it generates (Figure \ref{fig:vizinterface}). While plots
are still realized in some cases despite this error (Vega-Lite incorporates
good defaults), the affected axis is not interpretable. Another limitation of
the model is observed in selecting fields (attributes) 
of the input data to visualize --- the model sometime selects fields that are unintuitive or have little
information value. For example, a frequency plot of grouped longitude and
latitude values does not provide much information. Finally, the model generates
relatively simple visualizations --- univariate plots (which can serve as data
field summaries) and bivariate plots. It is unable to apply complex transforms,
use  multiple variables. 

\bfparhead{Training Data} 
While further experimentation is required, our intuition is that the
limitations mentioned above reflect limitations in both the size and diversity of our training data. Our goal with Data2Vis was to evaluate the viability of machine translation in generating valid visualization specifications, we have conducted our experiments with a relatively small dataset (4300 examples up sampled to 215,000 training pairs). While  our current results provide insights, we believe a larger and more diversified training dataset will improve learning and model generalization. Another limitation with our training data is related to our training pair generation strategy. Currently, we construct our source tokens from a single row from a dataset which is then preprocessed before training. While this approach shortens the input sequence length, a requirement for us to efficiently train our model, the model can only learn properties of each  field (e.g. length, content, numeric type, string type) as opposed to properties of the distribution of the field (e.g mean, range, categories etc.) which encode useful signals for data visualization.
 
\subsection{Future Work}


\bfparhead{Eliciting More Training Data} 
Naturally, addressing limitations with our training data constitutes the next
step for future work. We plan to conduct a structured data collection aimed at
generating visualization examples across a large number of datasets,
visualization types (bar, point, area, chart etc), transforms, complexity
(number of variables), interactions and visualization languages.  We will also
explore strategies to improve the training process that guide the model towards
learning properties of the distribution for a given field.

\bfparhead{Extending Data2Vis to Generate Multiple Plausible Visualizations}
Data2Vis is currently implemented as a sequence to sequence translation model.
Sequence models work very well for domains where it is desirable to have fixed
mappings of input sequences to output sequences (text summarization, image
captioning, language translation, etc). It is generally expected that a
sentence in one language always maps to the same sentence in a different
language, and acceptable if a passage always maps to the same summary or an
image to the same caption. However, when applied to the task of data
visualization, it is desirable that input data maps to  \textit{multiple} valid
visualizations. In the current work, we address this by exploiting beam search decoding to generate multiple visualizations based on a single dataset. A related avenue for future work is to explore generative models that can learn a probability distribution of effective visualizations, enabling \textit{one to many sequence} mappings between data and visualization specifications through sampling. 

\bfparhead{Targeting Additional Grammars} Building on results from Data2Vis,
important next steps also include efforts to train models that can map input
data to multiple different visualization specification languages, including
ggplot2,  given a dataset.  This line of research may also explore training
models that learn direct mappings between different visualization specification
languages, enabling visualization specification reuse across languages and platforms.

\bfparhead{Natural Language and Visualization Specification}
We propose the exploration of models that generate visualizations conditioned
on natural language text in addition to datasets. A potential approach is to
first explore how users might describe or express visualizations for a given
dataset and use this knowledge in generation of triplets---natural language
description, data sequence, and visualization specification. These data points
can then be leveraged in training a model that learns to generate
visualizations based on natural language descriptions. These models can extend
the expressive capabilities of existing systems that integrate multimodal
interactions and visualizations for exploring data. Conversely, we can use
textual descriptions of visualizations to automatically generate captions for
them, akin to image caption generation (e.g.,
~\cite{karpathy2015deep,vinyals2015show,xu2015show}). 


\section{Conclusion}
The history of data visualization is rich with work that  treats visualization
from a linguistic perspective.  Bertin systematized  data visualization as ``a
language  for the eye''~\cite{Bertin_1983}. Adopting Bertin's analogy,
Mackinlay~\cite{Mackinlay_1986} viewed visualizations as sentences of a
graphical language and formalized a model based on ``expressiveness'' and
``effectiveness'' criteria, borrowing concepts from formal languages.
Subsequent research also introduced various ``grammars'' of visualization
specification. 

We significantly extend this earlier perspective and formulate data
visualization as a sequence to sequence translation problem where we translate
data specifications to visualization specifications. We train a deep sequence
to sequence model and demonstrate its efficacy generating univariate and
bivariate plots. We also identify initial failure conditions, offer ideas for
their remediation and an agenda for future work.

It is our belief that the problem formulation and model presented in this work
represents an appropriate baseline for future work in automated generation of
visualizations using deep learning approaches. Our approach sets the stage for
systems that learn to generate visualizations at scale with implications for
the development of guided visual data exploration systems.

\section{Acknowledgments}
We thank Jorge Poco for making the Vega-Lite 
corpus~\cite{2017-reverse-engineering-vis} available.

\bibliographystyle{abbrv-doi}

\bibliography{paper}

\begin{thebibliography}{10}

\bibitem{maceachren:book95}
M.~M. Alan.
\newblock {\em How Maps Work: Representation, Visualization, and Design}.
\newblock Guilford Press, 1995.

\bibitem{Asimov_1985}
D.~Asimov.
\newblock The grand tour: A tool for viewing multidimensional data.
\newblock {\em SIAM J. Sci. Stat. Comput.}, 6(1), 1985.

\bibitem{Bahdanau2014}
D.~Bahdanau, K.~Cho, and Y.~Bengio.
\newblock {Neural Machine Translation by Jointly Learning to Align and
  Translate}.
\newblock {\em ICLR 2015}, sep 2014.

\bibitem{BalogGBNT16}
M.~Balog, A.~L. Gaunt, M.~Brockschmidt, S.~Nowozin, and D.~Tarlow.
\newblock Deepcoder: Learning to write programs.
\newblock {\em CoRR}, abs/1611.01989, 2016.

\bibitem{shortridge:ac82}
S.~B. Barbara.
\newblock Stimulus processing models from psychology: can we use them in
  cartography?
\newblock {\em The American Cartographer}, 9:155--167, 1982.

\bibitem{beltramelli2017pix2code}
T.~Beltramelli.
\newblock pix2code: Generating code from a graphical user interface screenshot.
\newblock {\em arXiv preprint arXiv:1705.07962}, 2017.

\bibitem{Bertin_1983}
J.~Bertin.
\newblock {\em Semiology of Graphics}.
\newblock University of Wisconsin Press, 1983.

\bibitem{BojanowskiJM15}
P.~Bojanowski, A.~Joulin, and T.~Mikolov.
\newblock Alternative structures for character-level rnns.
\newblock {\em CoRR}, abs/1511.06303, 2015.

\bibitem{bostock2009protovis}
M.~Bostock and J.~Heer.
\newblock Protovis: A graphical toolkit for visualization.
\newblock {\em IEEE TVCG (Proc. InfoVis)}, 2009.

\bibitem{bostock2011d3}
M.~Bostock, V.~Ogievetsky, and J.~Heer.
\newblock D3: Data-driven documents.
\newblock {\em IEEE TVCG (Proc. InfoVis)}, 2011.

\bibitem{Bouali_2016}
F.~Bouali, A.~Guettala, and G.~Venturini.
\newblock {VizAssist}: an interactive user assistant for visual data mining.
\newblock {\em Vis. Comput.}, 32(11):1447--1463, 2016.

\bibitem{boulanger2012modeling}
N.~Boulanger-Lewandowski, Y.~Bengio, and P.~Vincent.
\newblock Modeling temporal dependencies in high-dimensional sequences:
  Application to polyphonic music generation and transcription.
\newblock {\em arXiv:1206.6392}, 2012.

\bibitem{britz2017architectures}
D.~{Britz}, A.~{Goldie}, T.~{Luong}, and Q.~{Le}.
\newblock {Massive Exploration of Neural Machine Translation Architectures}.
\newblock {\em ArXiv e-prints}, Mar. 2017.

\bibitem{casner:tog91}
S.~M. Casner.
\newblock Task-analytic approach to the automated design of graphic
  presentations.
\newblock {\em ACM Trans. Graphics}, 10(2):111--151, 1991.

\bibitem{ChenLS18}
X.~Chen, C.~Liu, and D.~Song.
\newblock Tree-to-tree neural networks for program translation.
\newblock {\em CoRR}, abs/1802.03691, 2018.

\bibitem{ChoMGBSB14}
K.~Cho, B.~van Merrienboer, {\c{C}}.~G{\"{u}}l{\c{c}}ehre, F.~Bougares,
  H.~Schwenk, and Y.~Bengio.
\newblock Learning phrase representations using {RNN} encoder-decoder for
  statistical machine translation.
\newblock {\em CoRR}, abs/1406.1078, 2014.

\bibitem{cleveland:jasa84}
W.~S. Cleveland and R.~McGill.
\newblock Graphical perception: Theory, experimentation, and application to the
  development of graphical methods.
\newblock {\em J. Amer. Statist. Assoc.}, 79(387):531--554, 1984.

\bibitem{Demiralp:2017:VLDB}
{\c{C}}.~Demiralp, P.~J. Haas, S.~Parthasarathy, and T.~Pedapati.
\newblock Foresight: Recommending visual insights.
\newblock {\em Proc. VLDB Endow.}, 10(12):1937--1940, 2017.

\bibitem{Demiralp_2014a}
{\c{C}}.~Demiralp, C.~Scheidegger, G.~Kindlmann, D.~Laidlaw, and J.~Heer.
\newblock Visual embedding: A model for visualization.
\newblock {\em IEEE CG\&A}, 2014.

\bibitem{DevlinUBSMK17}
J.~Devlin, J.~Uesato, S.~Bhupatiraju, R.~Singh, A.~Mohamed, and P.~Kohli.
\newblock Robustfill: Neural program learning under noisy {I/O}.
\newblock {\em CoRR}, abs/1703.07469, 2017.

\bibitem{Data2Vis:2018:GitHub}
V.~Dibia.
\newblock Data2vis: Automatic generation of data visualizations using sequence
  to sequence recurrent neural networks.
\newblock \url{https://github.com/victordibia/data2vis}, 2017.

\bibitem{Dibia:2018:Data2Vis}
V.~Dibia and {\c{C}}.~Demiralp.
\newblock {Data2Vis}: Automatic generation of visualizations using
  sequence-to-sequence recurrent neural networks, 2018.

\bibitem{Dong2016LanguageAttention}
L.~Dong and M.~Lapata.
\newblock {Language to Logical Form with Neural Attention}.
\newblock jan 2016.

\bibitem{eck2002first}
D.~Eck and J.~Schmidhuber.
\newblock A first look at music composition using lstm recurrent neural
  networks.
\newblock {\em Istituto Dalle Molle Di Studi Sull Intelligenza Artificiale},
  103, 2002.

\bibitem{PRIM9_1974}
M.~A. Fisherkeller, J.~H. Friedman, and J.~W. Tukey.
\newblock Prim-9: An interactive multidimensional data display and analysis
  system.
\newblock In {\em Proc. Fourth International Congress for Stereology}, 1974.

\bibitem{learningto99gers}
F.~A. Gers, J.~Schmidhuber, and F.~Cummins.
\newblock Learning to forget: Continual prediction with lstm.
\newblock {\em Neural Computation}, 12:2451--2471, 1999.

\bibitem{Gotz_2008}
D.~Gotz and Z.~Wen.
\newblock Behavior-driven visualization recommendation.
\newblock In {\em ACM IUI}, pp. 315--324, 2009.

\bibitem{Graves2012}
A.~Graves.
\newblock {Sequence Transduction with Recurrent Neural Networks}, nov 2012.

\bibitem{HaSketchRNNE17}
D.~Ha and D.~Eck.
\newblock A neural representation of sketch drawings.
\newblock {\em CoRR}, abs/1704.03477, 2017.

\bibitem{hanrahan2006vizql}
P.~Hanrahan.
\newblock Vizql: a language for query, analysis and visualization.
\newblock In {\em Proceedings of the 2006 ACM SIGMOD international conference
  on Management of data}, pp. 721--721. ACM, 2006.

\bibitem{lstm97hochreiter}
S.~Hochreiter and J.~Schmidhuber.
\newblock Long short-term memory.
\newblock 9:1735--80, 12 1997.

\bibitem{Hu:2018:VizML}
K.~Z. Hu, M.~A. Bakker, S.~Li, T.~Kraska, and C.~A. Hidalgo.
\newblock Vizml: {A} machine learning approach to visualization recommendation,
  2018.

\bibitem{jean2014using}
S.~Jean, K.~Cho, R.~Memisevic, and Y.~Bengio.
\newblock On using very large target vocabulary for neural machine translation.
\newblock {\em arXiv:1412.2007}, 2014.

\bibitem{johnson2017generating}
D.~D. Johnson.
\newblock Generating polyphonic music using tied parallel networks.
\newblock In {\em International Conference on Evolutionary and Biologically
  Inspired Music and Art}, pp. 128--143. Springer, 2017.

\bibitem{kalchbrenner2013recurrent}
N.~Kalchbrenner and P.~Blunsom.
\newblock Recurrent continuous translation models.
\newblock In {\em Proceedings of the 2013 Conference on Empirical Methods in
  Natural Language Processing}, pp. 1700--1709, 2013.

\bibitem{karpathy2015deep}
A.~Karpathy and L.~Fei-Fei.
\newblock Deep visual-semantic alignments for generating image descriptions.
\newblock In {\em Proc. IEEE CVPR}, pp. 3128--3137, 2015.

\bibitem{KarpathyJL15}
A.~Karpathy, J.~Johnson, and F.~Li.
\newblock Visualizing and understanding recurrent networks.
\newblock {\em CoRR}, abs/1506.02078, 2015.

\bibitem{DBLP:journals/corr/abs-1710-10196}
T.~Karras, T.~Aila, S.~Laine, and J.~Lehtinen.
\newblock Progressive growing of gans for improved quality, stability, and
  variation.
\newblock {\em CoRR}, abs/1710.10196, 2017.

\bibitem{Key_2012}
A.~Key, B.~Howe, D.~Perry, and C.~Aragon.
\newblock {VizDeck}.
\newblock In {\em ACM SIGMOD}, vol. 681--684, 2012.

\bibitem{avd2014infovis}
G.~Kindlmann and C.~Scheidegger.
\newblock An algebraic process for visualization design.
\newblock {\em IEEE TVCG}, 20:2181--2190, 2014.

\bibitem{lewandowsky:jasa89}
S.~Lewandowsky and I.~Spence.
\newblock Discriminating strata in scatterplots.
\newblock {\em Journal of American Statistical Association}, 84(407):682--688,
  1989.

\bibitem{LingGHKSWB16}
W.~Ling, E.~Grefenstette, K.~M. Hermann, T.~Kocisk{\'{y}}, A.~Senior, F.~Wang,
  and P.~Blunsom.
\newblock Latent predictor networks for code generation.
\newblock {\em CoRR}, abs/1603.06744, 2016.

\bibitem{Yuyu:2018:DeepEye}
Y.~Luo, X.~Qin, N.~Tang, G.~Li, and X.~Wang.
\newblock Deepeye: Creating good data visualizations by keyword search.
\newblock In {\em Proceedings of the 2018 International Conference on
  Management of Data}, SIGMOD, pp. 1733--1736, 2018.

\bibitem{Luong2016}
M.-T. Luong and C.~D. Manning.
\newblock {Achieving Open Vocabulary Neural Machine Translation with Hybrid
  Word-Character Models}.
\newblock {\em ACL 2016}, apr 2016.

\bibitem{Luong2015}
M.-T. Luong, H.~Pham, and C.~D. Manning.
\newblock {Effective Approaches to Attention-based Neural Machine Translation}.
\newblock {\em EMNLP 2015}, aug 2015.

\bibitem{luong2014addressing}
M.-T. Luong, I.~Sutskever, Q.~V. Le, O.~Vinyals, and W.~Zaremba.
\newblock Addressing the rare word problem in neural machine translation.
\newblock {\em arXiv:1410.8206}, 2014.

\bibitem{Mackinlay_1986}
J.~Mackinlay.
\newblock Automating the design of graphical presentations of relational
  information.
\newblock {\em {ACM} Trans. Graphics}, 5(2):110--141, 1986.

\bibitem{showme:infovis07}
J.~Mackinlay, P.~Hanrahan, and C.~Stolte.
\newblock Show me: Automatic presentation for visual analysis.
\newblock {\em IEEE TVCG}, 13(6):1137--1144, 2007.

\bibitem{Moritz:2019:Draco}
D.~Moritz, C.~Wang, G.~Nelson, H.~Lin, A.~M. Smith, B.~Howe, and J.~Heer.
\newblock Formalizing visualization design knowledge as constraints: Actionable
  and extensible models in draco.
\newblock {\em IEEE TVCG (Proc. InfoVis)}, 2019.

\bibitem{Nallapati2016}
R.~Nallapati, B.~Zhou, C.~N. dos Santos, C.~Gulcehre, and B.~Xiang.
\newblock {Abstractive Text Summarization Using Sequence-to-Sequence RNNs and
  Beyond}.
\newblock {\em The SIGNLL Conference on Computational Natural Language Learning
  (CoNLL), 2016}, feb 2016.

\bibitem{ParisottoMSLZK16}
E.~Parisotto, A.~Mohamed, R.~Singh, L.~Li, D.~Zhou, and P.~Kohli.
\newblock Neuro-symbolic program synthesis.
\newblock {\em CoRR}, abs/1611.01855, 2016.

\bibitem{2017-reverse-engineering-vis}
J.~Poco and J.~Heer.
\newblock Reverse-engineering visualizations: Recovering visual encodings from
  chart images.
\newblock {\em Computer Graphics Forum (Proc. EuroVis)}, 2017.

\bibitem{reed2016generative}
S.~Reed, Z.~Akata, X.~Yan, L.~Logeswaran, B.~Schiele, and H.~Lee.
\newblock Generative adversarial text to image synthesis.
\newblock {\em arXiv preprint arXiv:1605.05396}, 2016.

\bibitem{Roth_1994}
S.~F. Roth, J.~Kolojechick, J.~Mattis, and J.~Goldstein.
\newblock Interactive graphic design using automatic presentation knowledge.
\newblock In {\em ACM Human Factors in Computing Systems (CHI)}, 1994.

\bibitem{Saket:2018:TVCG}
B.~Saket, A.~Endert, and {\c{C}}.~Demiralp.
\newblock Task-based effectiveness of basic visualizations.
\newblock {\em IEEE TVCG}, 2018.

\bibitem{Saket:2018:Learning}
B.~Saket, D.~Moritz, H.~Lin, V.~Dibia, {\c{C}}.~Demiralp, and J.~Heer.
\newblock Beyond heuristics: Learning visualization design, 2018.

\bibitem{satyanarayan2017vegalite}
A.~Satyanarayan, D.~Moritz, K.~Wongsuphasawat, and J.~Heer.
\newblock Vega-lite: A grammar of interactive graphics.
\newblock {\em IEEE TVCG (Proc. InfoVis)}, 2017.

\bibitem{satyanarayan2016vega}
A.~Satyanarayan, R.~Russell, J.~Hoffswell, and J.~Heer.
\newblock Reactive vega: A streaming dataflow architecture for declarative
  interactive visualization.
\newblock {\em IEEE TVCG (Proc. InfoVis)}, 2016.

\bibitem{Schuster1997}
M.~Schuster and K.~K. Paliwal.
\newblock {Bidirectional recurrent neural networks}.
\newblock {\em IEEE Transactions on Signal Processing}, 45(11):2673--2681,
  1997. doi: {{%
10\hspace{.1pt}\discretionary{.}{%
}{.}\hspace{.4pt}1109\discretionary{/}{%
}{/}78\hspace{.1pt}\discretionary{.}{%
}{.}\hspace{.4pt}650093}}


\bibitem{sennrich2016edinburgh}
R.~Sennrich, B.~Haddow, and A.~Birch.
\newblock Edinburgh neural machine translation systems for wmt 16.
\newblock {\em arXiv preprint arXiv:1606.02891}, 2016.

\bibitem{seo:infovis04}
J.~Seo and B.~Shneiderman.
\newblock A rank-by-feature framework for unsupervised multidimensional data
  exploration using low dimensional projections.
\newblock In {\em Procs. InfoVis}, pp. 65--72, 2004.

\bibitem{shepard:sci87}
R.~N. Shepard.
\newblock Toward a universal law of generalization for psychological science.
\newblock {\em Science}, 237:1317--1323, 1987.

\bibitem{Siddiqui_2016}
T.~Siddiqui, A.~Kim, J.~Lee, K.~Karahalios, and A.~Parameswaran.
\newblock Effortless data exploration with {Zenvisage}.
\newblock {\em PVLDB}, 10(4):457--468, 2016.

\bibitem{Stolte_2002}
C.~Stolte, D.~Tang, and P.~Hanrahan.
\newblock Polaris: a system for query, analysis, and visualization of
  multidimensional relational databases.
\newblock {\em {IEEE TVCG}}, 8(1):52--65, 2002. doi: {{%
10\hspace{.1pt}\discretionary{.}{%
}{.}\hspace{.4pt}1109\discretionary{/}{%
}{/}2945\hspace{.1pt}\discretionary{.}{%
}{.}\hspace{.4pt}981851}}


\bibitem{SumitChopra2016}
{Sumit Chopra}, {Michael Auli}, and {Alexander M. Rush}.
\newblock {Abstractive sentence summarization with attentive recurrent neural
  networks}.
\newblock In {\em Proceedings of NAACL-HLT}, 2016.

\bibitem{SutskeverVL14}
I.~Sutskever, O.~Vinyals, and Q.~V. Le.
\newblock Sequence to sequence learning with neural networks.
\newblock {\em CoRR}, abs/1409.3215, 2014.

\bibitem{topk2017sigmod}
B.~Tang, S.~Han, M.~L. Yiu, R.~Ding, and D.~Zhang.
\newblock Extracting top-k insights from multi-dimensional data.
\newblock In {\em ACM SIGMOD}, pp. 1509--1524, 2017.

\bibitem{Vartak_2015b}
M.~Vartak, S.~Huang, T.~Siddiqui, S.~Madden, and A.~Parameswaran.
\newblock Towards visualization recommendation systems.
\newblock In {\em DSIA Workshop}, 2015.

\bibitem{Vartak_2015a}
M.~Vartak, S.~Rahman, S.~Madden, A.~Parameswaran, and N.~Polyzotis.
\newblock See{DB}: Efficient data-driven visualization recommendations to
  support visual analytics.
\newblock {\em PVLDB}, 8(13):2182--2193, 2015.

\bibitem{vinyals2015show}
O.~Vinyals, A.~Toshev, S.~Bengio, and D.~Erhan.
\newblock Show and tell: A neural image caption generator.
\newblock In {\em Computer Vision and Pattern Recognition (CVPR), 2015 IEEE
  Conference on}, pp. 3156--3164. IEEE, 2015.

\bibitem{wickham2010layered}
H.~Wickham.
\newblock A layered grammar of graphics.
\newblock {\em Journal of Computational and Graphical Statistics}, 19(1):3--28,
  2010.

\bibitem{wilkinson:book99}
L.~Wilkinson.
\newblock {\em The Grammar of Graphics}.
\newblock Springer, 1st ed., 1999.

\bibitem{scagnostics2005infovis}
L.~Wilkinson, A.~Anand, and R.~Grossman.
\newblock Graph-theoretic scagnostics.
\newblock In {\em Proc. InfoVis}, pp. 157--164, 2005.

\bibitem{wills2017brunel}
G.~Wills.
\newblock Brunel v2.5.
\newblock https://github.com/Brunel-Visualization/Brunel, 2017.
\newblock Accessed: 2018-04-04.

\bibitem{Wills_2008}
G.~Wills and L.~Wilkinson.
\newblock {AutoVis}: Automatic visualization.
\newblock {\em Info. Visual.}, 9(1):47--69, 2008.

\bibitem{Wongsuphasawat_2016CompassQL}
K.~Wongsuphasawat, D.~Moritz, A.~Anand, J.~Mackinlay, B.~Howe, and J.~Heer.
\newblock {Towards a general-purpose query language for visualization
  recommendation}.
\newblock In {\em Procs. HILDA}, 2016.

\bibitem{Wongsuphasawat_2016}
K.~Wongsuphasawat, D.~Moritz, A.~Anand, J.~Mackinlay, B.~Howe, and J.~Heer.
\newblock Voyager: Exploratory analysis via faceted browsing of visualization
  recommendations.
\newblock {\em IEEE TVCG}, 22(1):649--658, 2016.

\bibitem{Wongsuphasawat_2017}
K.~Wongsuphasawat, Z.~Qu, D.~Moritz, R.~Chang, F.~Ouk, A.~Anand, J.~Mackinlay,
  B.~Howe, and J.~Heer.
\newblock Voyager 2: Augmenting visual analysis with partial view
  specifications.
\newblock In {\em ACM CHI}, 2017.

\bibitem{wu2016google}
Y.~Wu, M.~Schuster, Z.~Chen, Q.~V. Le, M.~Norouzi, W.~Macherey, M.~Krikun,
  Y.~Cao, Q.~Gao, K.~Macherey, et~al.
\newblock Google's neural machine translation system: Bridging the gap between
  human and machine translation.
\newblock {\em arXiv:1609.08144}, 2016.

\bibitem{xu2015show}
K.~Xu, J.~Ba, R.~Kiros, K.~Cho, A.~Courville, R.~Salakhudinov, R.~Zemel, and
  Y.~Bengio.
\newblock Show, attend and tell: Neural image caption generation with visual
  attention.
\newblock In {\em Proc. ICML}, pp. 2048--2057, 2015.

\bibitem{YinN17}
P.~Yin and G.~Neubig.
\newblock A syntactic neural model for general-purpose code generation.
\newblock {\em CoRR}, abs/1704.01696, 2017.

\bibitem{Zhong2017Seq2SQLLearning}
V.~Zhong, C.~Xiong, and R.~Socher.
\newblock {Seq2SQL: Generating Structured Queries from Natural Language using
  Reinforcement Learning}.
\newblock aug 2017.

\end{thebibliography}

\begin{figure*}[ht] 
  \includegraphics[width=\textwidth]{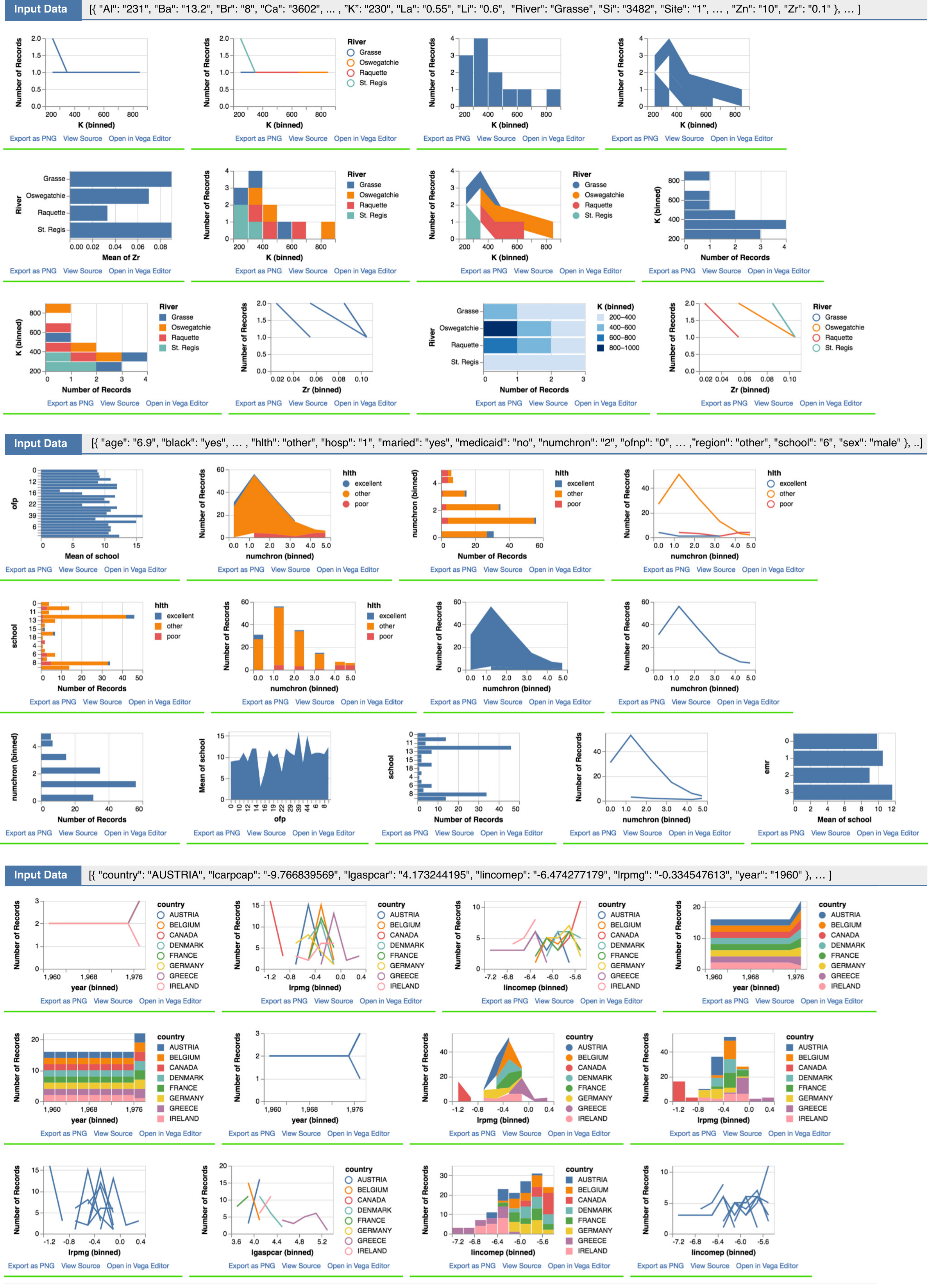}
  \caption{Examples of visualizations generated with beam search.\label{fig:examples}} 
\end{figure*}

\end{document}